\documentclass[journal=jacsat,manuscript=article]{achemso}

\usepackage[version=3]{mhchem} 

\author{Jérôme TRIBOLLET}
\email{tribollet@unistra.fr}
\affiliation[Strasbourg University]
{Institut de Chimie de Strasbourg, Strasbourg University, UMR 7177 (CNRS-UDS),\\
4 rue Blaise Pascal, CS 90032, F-67081 Strasbourg Cedex, 
France}

\title[An \textsf{achemso} demo]
  {Hybrid paramagnetic-ferromagnetic\\ quantum computer design based on\\ electron spin arrays and a ferromagnetic nanostripe}

\abbreviations{EPR (Electron Paramagnetic Resonance)} 
  

\begin{document}
\begin{abstract}
Designing an assembly of quantum nano-objects which can interact between themselves and be manipulated by external fields, while staying isolated from their noisy environment is the key for the development of future quantum technologies, such as quantum computers and sensors. Electron spins placed in a magnetic field gradient, interacting by dipolar magnetic couplings  and manipulated by microwave pulses represent a possible architecture for a quantum computer. Here, a general design for the practical implementation of such nanodevice is presented and illustrated on the example of electron spins in silicon carbide placed nearby a permalloy ferromagnetic nanostripe. Firstly, the confined spin wave resonance spectrum of the nanostripe and the properties of its magnetic field gradient are calculated. Then, one shows how to avoid microwave driven electron spin decoherence. Finally, one shows that decoherence due to ferromagnetic fluctuations is negligible below room temperature for spins placed far enough from the nanostripe.   \textsl{}
    
\end{abstract}

\section{Introduction}
During last decades, the development of nanotechnologies has allowed the experimental demonstration of entanglement between the quantum states of different pairs of nanosystems~\cite{Blatt2008,Neeley2010,Wrachtrup2010,Wrachtrup2013}, a key preliminary step towards quantum information processing. However, up scaling those experiments to at least ten coupled  quantum systems, called quantum bits or qubits, on a single chip remains extremely challenging. This difficulty is linked to the need to couple quantum bits over nanoscale distances to perform conditional logical operations, except when some alternative strategy for long distance entanglement is available~\cite{Burkard2006,Hanson2013}. In the context of quantum bits encoded on electron spins in solids, the easiest solution to couple them is to use the dipolar magnetic coupling~\cite{DasSarma2004,Wrachtrup2010,Wrachtrup2013}, which has however two drawbacks. Firstly, it becomes too weak beyond ten nanometers for efficient quantum states entanglement. Secondly, it is a permanent coupling, which requires methods to effectively switch on/off those couplings. The up scaling problem is thus directly related in this context to the difficulty to precisely position the electron spins on the chip and also to manipulate them with nanoscale resolution. State of art nanolithography and ion implantation methods used to build silicon~\cite{Vrijen2000} or diamond~\cite{Awschalom2010,Wrachtrup2010,Wrachtrup2013} based small scale quantum devices with few coupled electron spins are still far from reaching the ten qubits scale. Building solid state arrays of dipolar coupled electron spin qubits thus requieres an alternative nanotechnology and also a method and a nanodevice allowing to switch on/off those dipolar couplings. Transmission Electron Microscope (TEM) based nanofabrication methods with nanometer scale precision were recently demonstrated with carbon nanotubes~\cite{Banhart2009} and graphene materials~\cite{WarnerTEM2013}, and  could provide the required alternative nanotechnology.  Also, few years ago, it was shown that selective on/off switching of the dipolar coupling is possible using sequences of microwave pulses applied to dipolar coupled electron spins submitted to a magnetic field gradient~\cite{DasSarma2004}. The problem was thus shifted to the production of strong magnetic field gradient.  In the field of quantum computing with nuclear spins, it was proposed to use dysprosium nanoferromagnet to produce a huge magnetic field gradient~\cite{Yamamoto2000}. However, in the context of electron spin qubits, the nanoferromagnet also produces unwanted additional decoherence of the electron spins~\cite{tribolletEPJBtheodeco} due either to incoherent thermal excitation or to coherent microwave excitation of the spin waves~\cite{Kittel1958} confined in the nanoferromagnet~\cite{Jorzick2002,Lee2010}. 
Here I present a quantum register design containing a permalloy ferromagnetic nanostripe and nearby arrays of electron spins  in silicon carbide (SiC). The spin qubits are silicon vacancies~\cite{Awschalom2013,Baranov2012,Itoh1990,Wimbauer1997,Lefevre2009} that could be created in SiC using TEM nanofabrication methods~\cite{Banhart2009,WarnerTEM2013,Steeds2002,Baranov2012,Lefevre2009}. Using magnetostatic theory~\cite{Yamamoto2000}, the  Landau Lifschitz equation of motion of magnetization in  the nanoferromagnet~\cite{Lee2010}, a one dimensional Schrodinger equation solved numerically by a transfer matrix method~\cite{Jirauschek2009}, and the density matrix theory of spin qubit decoherence~\cite{tribolletEPJBtheodeco}, I show how to design simultaneously the magnetic field gradient and the confined spin waves spectrum of the nanoferromagnet in order to avoid additional decoherence. The designed quantum register could be operated with microwave pulses~\cite{schweiger2001} and read out with high sensitivity using ensemble optical measurements of the paramagnetic resonance of silicon vacancy electron spins in SiC~\cite{Awschalom2013,Baranov2012}.\\\\  

\section{Design and fabrication of a nanodevice with electron spin arrays nearby a ferromagnetic nanostripe}

The general design proposed here for quantum information processing could be applied, in principle, to all kind of electron spin qubits like phosphorous donors in silicon~\cite{Vrijen2000,Tyryshkin2011}, nitrogen donors or NV centers in diamonds~\cite{Awschalom2010,Wrachtrup2010,Wrachtrup2013}, transition metals ions or shallow donors in Zinc Oxide~\cite{tribolletEPLFe,tribolletEPJBtheodeco,GamelinNCZnOMn,MortonpiezoZnOMn,KhalifEPLMnCo}, N@C60 molecules~\cite{HarneitNC60}, paramagnetic defects in graphene~\cite{Nair2012}, paramagnetic dangling bonds arrays on hydrogen passivated silicon surface~\cite{Bowler2013}, or supramolecular assemblies of paramagnetic molecules such as neutral radical molecules~\cite{Tamulis2007} or copper phtalocyanine molecules~\cite{Warner2013}, as long as regular dense arrays of them could be produced at precise positions nearby a ferromagnetic nanostripe. Without loss of generality and to be more quantitative, I focus here on a nanodevice made with silicon carbide and permalloy materials and build with top-down methods. The successive steps of two possible nanofabrication processes allowing to build such a nanodevice are described on figure 1a and 1b. The resulting nanodevice is shown on figure 1c.
\begin{figure} [ht]
\centering \includegraphics[width=0.50\textwidth]{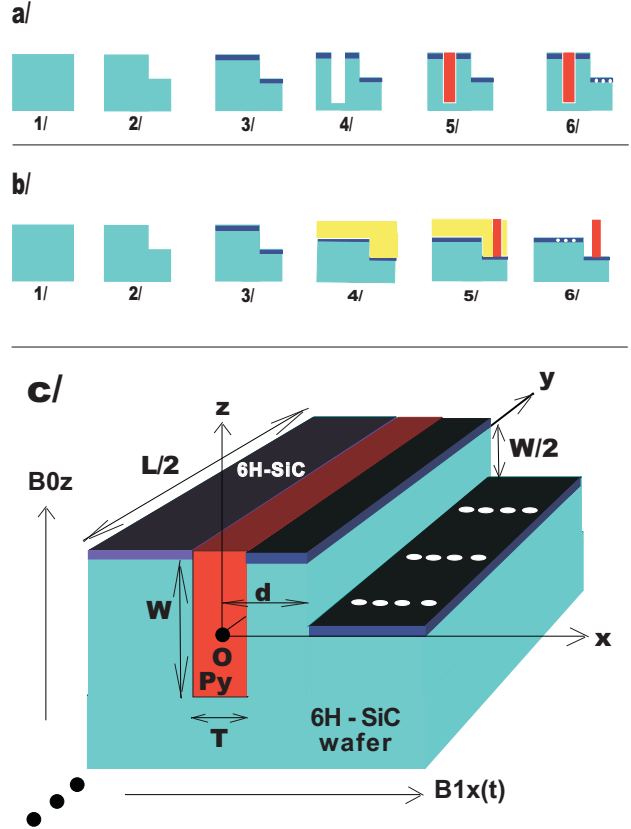}
\caption{\label{fig_01} a/ First process:  6H-SiC wafer / step in wafer produced by focused ion beam (FIB) milling/ epitaxial growth of isotopically purified (without nuclear spins) 6H-SiC on patterned wafer/ nanotrench created by focused ion beam (FIB) milling method~\cite{Menzel1998} nearby the edge of the step: depth $W$, width $T$, length $L$ / ion assisted trench filling methods~\cite{Cheryl1999} used to fill the nanotrench with Permalloy / TEM production~\cite{Banhart2009,Warner2013,Steeds2002,Baranov2012,Lefevre2009} of identical and parallel linear arrays of silicon vacancies (white dots) aligned along the x axis. b/ Second process: it combines electron beam lithography and permalloy plating~\cite{Yang2003}: first steps as in 1a and then: 1b4: resist deposition and planarization, 1b5: nanotrench produced in resist by e beam lithography, further filled with Permalloy (red), 1b6: removal of the resist, and then TEM production of silicon vacancies. c/ Design of the proposed SiC-Py quantum register corresponding to nanofabrication method a/. In the following calculations: $L\:=\:100\;\mu\:m$, $T\:=\:100\;n\:m$ , $W\:=\:800\;n\:m$.} 
\end{figure} 
Each of the three SiC polytypes, 4H, 6H, and 3C,  has some paramagnetic defects related to silicon vacancies which could potentially encode a good electron spin quantum bit~\cite{Awschalom2013,Baranov2012,Itoh1990,Wimbauer1997,Lefevre2009}. Some silicon vacancies defects in 6H-SiC has a spin S=3/2 with an isotropic g factor $g\:=\:2.0032$ and a small zero field splitting~\cite{Baranov2012}. Their spin state can also be prepared by optical alignment~\cite{Baranov2012} and it can be readout by the highly sensitive ODMR method (optically detected magnetic resonance)~\cite{Baranov2013}.  It seems thus highly relevant to encode an electron spin qubit on this kind of silicon vacancy defect in 6H-SiC (for example at X band or at higher microwave frequency, using: $\left|S=\:\frac{3}{2}\:,\:M_{S}=\:-\frac{3}{2}\:\right\rangle\:\equiv\:\left|0\right\rangle$ and $\left|S=\:\frac{3}{2}\:,\:M_{S}=\:-\frac{1}{2}\:\right\rangle\:\equiv\:\left|1\right\rangle$). In the following calculations, one simplifies the spin qubit model by using a simple pseudo spin s=1/2 with a g factor g=2, without any loss of generality.

\section{Dipolar magnetic field gradient outside the nanoferromagnet and effective Ising coupling between electron spin qubits}

The practical implementation of single and two qubits quantum gates on chains of dipolar coupled electron spin qubits, as those shown on figure 1, requires appropriate microwave pulse sequences and a strong magnetic field gradient along the direction of the spin chains of the nanodevice to transform their dipolar coupling into an effective Ising like coupling~\cite{DasSarma2004}. As here I also consider the possibility to perform ensemble measurements of the spin states of the spin qubits, for example using the highly sensitive ODMR method~\cite{Baranov2013}, the magnetic field gradient produced by the permalloy ferromagnetic stripe shown on figure 1 has not only to be strong enough along x, but also to be enough homogeneous along y and z. Strong enough along x means here that the energy difference between one qubit j (at position $x_{j}$) and the next one j+1 (at position $x_{j\:+\:1}$) along the spin chain has to be much larger than the energy associated to the dipolar coupling between them, which depends on their relative distance $l_{inter}$. In the nanodevice described on figure 1, $l_{inter}$ is thus also the periodicity of each spin chain. The condition for an effective Ising coupling translates into $\left|g\:\mu_{B}\:\left(B_{0}\:+\:B_{dip,z}\left(x_{j\:+\:1}\right)\right)\:-\:g\:\mu_{B}\:\left(B_{0}\:+\:B_{dip,z}\left(x_{j}\right)\right)\:\right|\:>>\:\frac{\mu_{0}\:\mu^{2}_{B}\:g^{2}}{4\:\pi\:l^{3}_{inter}}$, or approximately, $\left|g\:\mu_{B}\:\frac{d\:B_{dip,z}\left(x\right)}{d\:x}\:l_{inter}\right|\:>>\:\frac{\mu_{0}\:\mu^{2}_{B}\:g^{2}}{4\:\pi\:l^{3}_{inter}}$, where $B_{dip,z}\left(x\right)$ is the value of the z component of the dipolar magnetic field produced by the ferromagnetic nanostripe at position x, assuming that $z\:\approx\:0$ for each spin qubit, and that the y coordinate of each spin qubit does not matter for the value of $B_{dip,z}$ applied to the qubit. The irrelevance of the y coordinate is due to the approximate invariance by translation of the gradient produced by the long ferromagnetic nanostripe (infinite stripe model). This approximation is valid as long as the spin chains are created far from the edges of the ferromagnetic stripe along the y direction. Assuming a permalloy ferromagnetic stripe with a length $L\:=\:100\:\mu\:m$ along the y axis, a width $T\:=\:100\:n\:m$ along the x axis, and a depth $W\:=\:800\:n\:m$ along the z axis, further assuming that the ferromagnetic stripe is fully magnetized along the z axis by the static applied magnetic field $B_{0}$, one can calculate using magnetostatic theory~\cite{Yamamoto2000} the properties of the dipolar magnetic field produced by this nanoferromagnet, as shown on figure 2.
 \begin{figure} [ht]
\centering \includegraphics[width=1.10\textwidth]{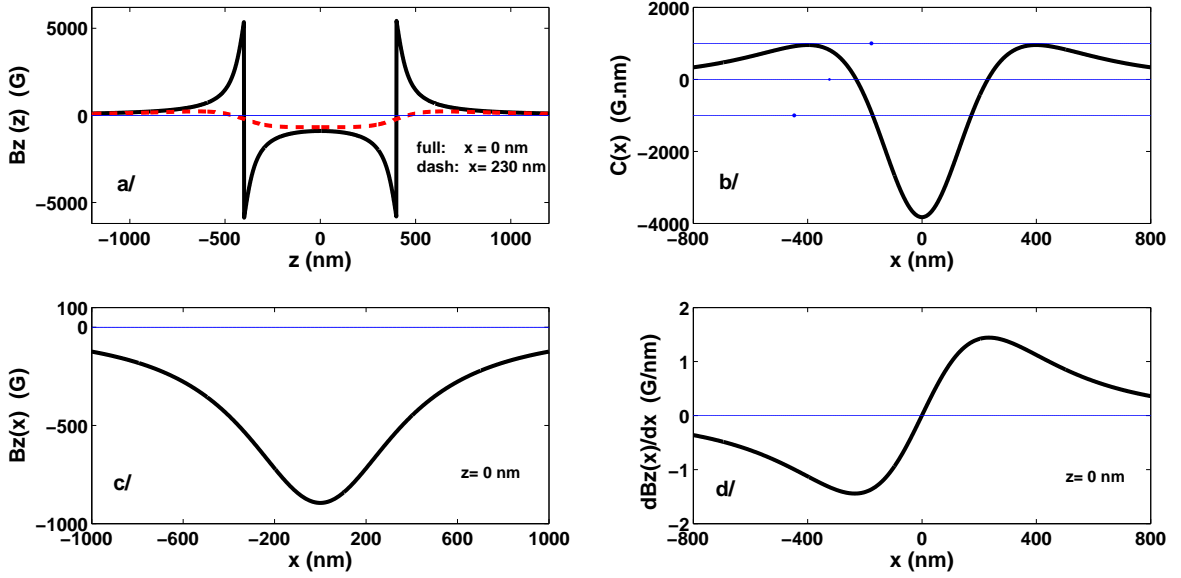}  
\caption{\label{fig_02} a/ $B_{dip,z}\left(z,\:x\right)$ for two different x position: x=0 (continuous black) and $x\:=\:x_{optim}\:=\:230\:n\:m$ (dash red). b/ $C\left(x\right)\:=\:\int^{+100}_{-100}\:dz\:\left(B_{dip,z}\left(x,\:z\right)\:-\:B_{dip,z}\left(x,\:0\right)\right)$. One defines the  position $x_{optim}$ by $C\left(x_{optim}\right)\:=\:0$. For the permalloy ferromagnetic nanostripe considered here (see figure 1), one finds $x_{optim}\:=\:230\:n\:m$. This position $x_{optim}$ is the position where the homogeneity of $B_{dip,z}\left(x,\:z\right)$ is optimal along the z direction. c/ $B_{dip,z}\left(z\:=\:0,\:x\right)$. d/ Gradient along x of the dipolar magnetic field produced by the ferromagnetic nanostripe, assuming z=0 and y=0: $\frac{d\:B_{dip,z}\left(x\right)}{dx}$. For the ferromagnetic Permalloy nanostripe, the saturation magnetization is $M_{sat,\:Py}$, with $\:\mu_{0}\:M_{sat,\:Py}\:=\:B_{sat,\:Py}\:=\:11300\:G$.}
\end{figure}

Figure 2 shows that as long as the spin qubits encoded onto silicon vacancies in SiC are created far from the edges of the ferromagnetic stripe along the y direction and close enough to the z=0 plane, then the relevant z component of the dipolar magnetic field produced by the ferromagnetic nanostripe can be considered as only dependent on x, and thus its gradient can also be approximately considered as being one dimensional along x. One finds here $B_{dip,z}\left(x_{optim}\:=\:230\:n\:m,\:0\right)\:\approx\:-\:676\:G$ and $\frac{d\:B_{dip,z}\left(x_{optim}\:=\:230\:n\:m,\:0\right)}{dx}\:\approx\:1.44\:G\:/\:nm$. From those results, one can thus estimate whether the condition given above to obtain an effective Ising coupling between the spin qubits is satisfied. Assuming that most of the spin qubits are created in SiC by TEM at x positions close to $x_{optim}\:=\:230\:n\:m$ and z=0, and further assuming a distance $l_{inter}\:=\:5\:nm$ between successive spin qubits along the x direction, one finds that $\left|g\:\mu_{B}\:\frac{d\:B_{dip,z}\left(x\right)}{d\:x}\:l_{inter}\right|$ is more than hundred times larger than the dipolar coupling energy $\frac{\mu_{0}\:\mu^{2}_{B}\:g^{2}}{4\:\pi\:l^{3}_{inter}}$. One could thus, at first sight, conclude that the proposed nanodevice described on figure 1, satisfies the requirements for the practical implementation of quantum information processing with dipolar coupled electron spin qubits manipulated by appropriate sequences of microwave pulses~\cite{DasSarma2004}. However, in order to avoid microwave driven decoherence, it is thus absolutely necessary to design the ferromagnetic nanostripe such that no spectral overlap between the precession frequencies of the spin qubits and the precession frequencies of the spin waves confined inside the ferromagnetic nanostripe~\cite{Kittel1958,Jorzick2002,Lee2010} occurs. 

\section{Designing the field sweep confined spin wave resonance spectrum to avoid coherently driven electron spin qubit decoherence}      

One assumes here that the permalloy ferromagnetic nanostripe is almost fully magnetized at equilibrium along z at sufficiently low temperature and sufficiently high magnetic field (see figure 1). The excitation of a ferromagnetic spin wave implies that the magnetization inside the ferromagnetic nanostripe is no more uniform. In other words that means that the dipolar magnetic field acting on the nearby electron spin qubits outside the nanoferromagnet will become fluctuating, leading to spin qubit decoherence~\cite{tribolletEPJBtheodeco,schweiger2001}. As it can be seen on figure 2a (in the case x=0) and on figure 3a, all the spins contained inside the ferromagnetic nanostripe produce a very inhomogeneous dipolar magnetic field inside the ferromagnetic nanostripe itself.
\begin{figure} [ht]
\centering \includegraphics[width=1.10\textwidth]{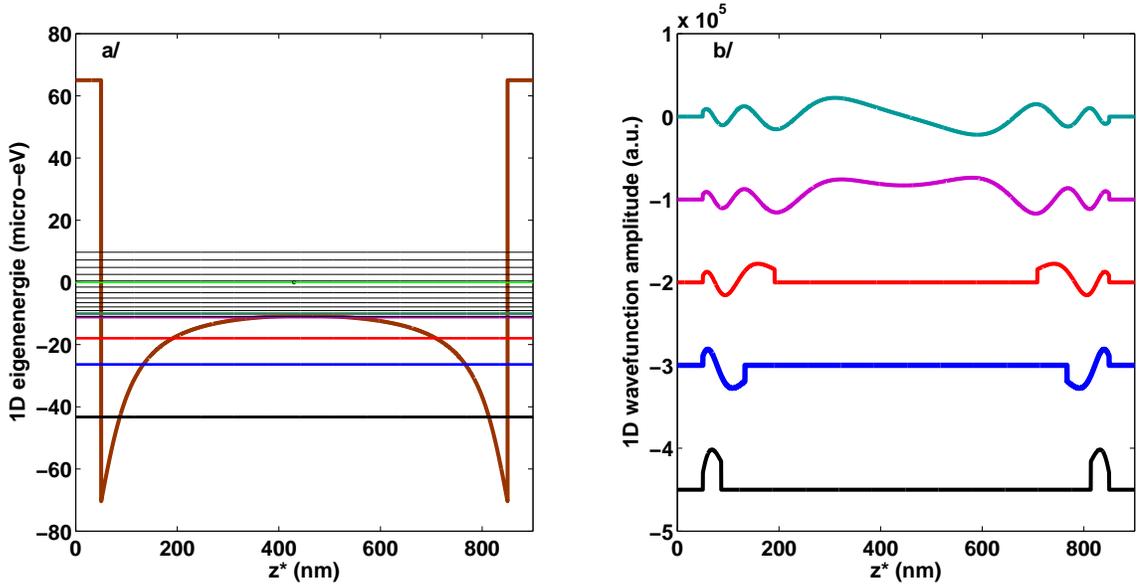} 
\caption{\label{fig_03} The one dimensional eigenenergies and eigenfunctions of the spin waves confined in the inhomogeneous effective potential existing inside the ferromagnetic nanostripe along the direction z of the static magnetic field applied: a/ one dimensional eigenenergies of the spin waves along z axis (horizontal lines) represented on top of the inhomogeneous effective confining potential inside the nanostripe; b/ one dimensional eigenfunctions of the spin waves confined along z (the out of equilibrium magnetization component $\delta\:m_{x}\left(z\right)$ has been plotted here). Here the permalloy ferromagnetic nanostripe has the dimensions: length $L\:=\:100\:\mu\:m$ along the y axis, width $T\:=\:100\:n\:m$ along the x axis, and depth $W\:=\:800\:n\:m$ along the z axis. It was further assumed that the ferromagnetic stripe is fully magnetized along the z axis by the static magnetic field applied $B_{0}$, and the magnetization of Permalloy at saturation was taken equal to $\mu_{0}\:M_{sat,\:Py}\:=\:B_{sat,\:Py}\:=\:11300\:G$. Note also that for numerical calculations, the new variable $z^{*}$ was used, defined by $z^{*}\:=\:z\:+\:450$ (in nm).} 
\end{figure} 
This leads to the confinement of the ferromagnetic spin waves~\cite{Jorzick2002} along the z axis, mainly on the edges of the ferromagnetic nanostripe. Here I have developed a method to calculate the confined spin wave resonance spectrum of a long ferromagnetic nanostripe fully magnetized. First, the  Landau Lifschitz equation of motion of magnetization in  the nanoferromagnet~\cite{Lee2010} is linearized. Then, the one dimensional eigenenergies and eigenfunctions of the spin waves confined in the inhomogeneous effective potential existing inside the ferromagnetic nanostripe along z are calculated, as it is shown respectively on figure 3a and figure 3b, using an effective Schrodinger equation solved numerically by a transfer matrix method~\cite{Jirauschek2009}. Then, the three dimensional eigenenergies of the confined spin waves are determined, and finally the resonance fields at which the microwave magnetic field of fixed frequency can excite a confined spin wave are determined, simply by writing the energy conservation rule. This rule says that the energy of the microwave photon absorbed should be equal to the three dimensional eigenenergie of the confined spin wave excited. It is important to note that the combination of the theoretical and of the experimental field sweep microwave absorption spectrum, provides an indirect method to determine the dipolar magnetic field produced by the ferromagnetic nanostripe on the spin qubits, because it is the same magnetization distribution that produces the dipolar magnetic field inside and outside the nanostripe. The full field sweep spectrum expected for the nanodevice described on figure 1 is plotted on figure 4a.
\begin{figure} [ht]
\centering \includegraphics[width=1.1\textwidth]{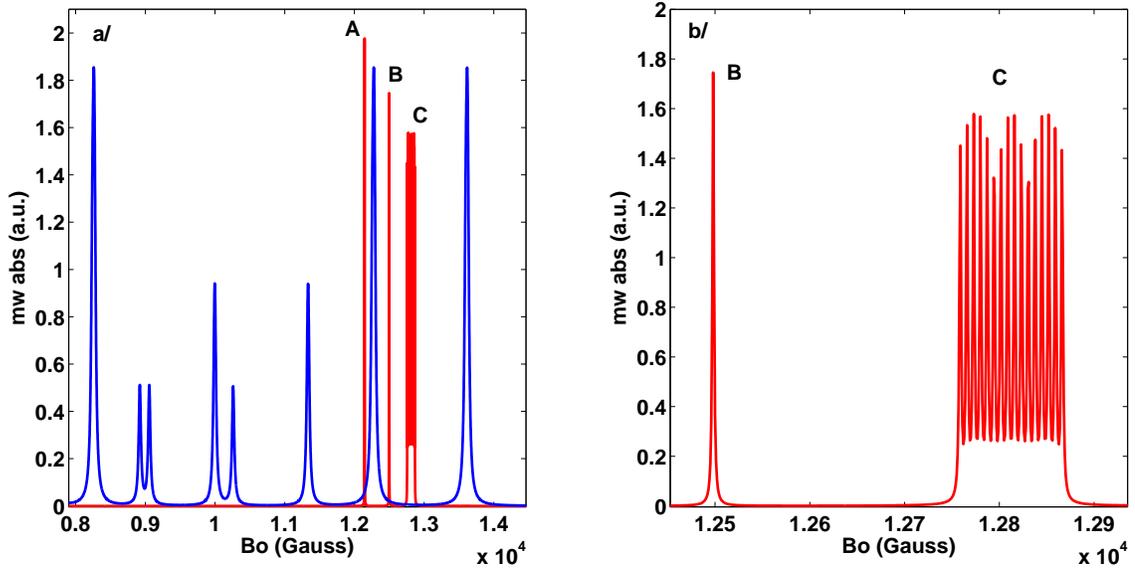} 
\caption{\label{fig_04} a/ Full field sweep microwave absorption spectrum of the nanodevice designed on figure 1, including the spin waves resonance spectrum (blue resonance lines) and the electron spin qubits paramagnetic resonance spectrum (red resonance lines), as it could be measured at a fixed microwave frequency $\nu\:=\:34\:GHz$ and with a varying magnetic field (see text for more details). b/ Zoom in the high magnetic field range of the field sweep microwave absorption spectrum of the proposed nanodevice, showing that no spectral overlap between the ferromagnetic spin waves and the electron spin qubits occur with this design (only red resonance lines are present in this field range). Note that the linewidth and the strength of the resonances (oscillator strength) were arbitrarily chosen here, that means, without taking into account the relative symmetry of the spin wave mode and of the microwave photon mode inside the microwave cavity, which determines the oscillator strength, in order to see any spin wave resonance satisfying the energy conservation rule alone.}
\end{figure}
It includes the field sweep spectrum expected for the ferromagnetic nanostripe (blue resonance lines with resonance magnetic fields occurring between 0.80 and 1.40 Tesla). It also includes the field sweep spectrum  expected for 16 electron spin qubits created at positions x close to $x_{optim}\:=\:230\:n\:m$, with x values comprised between $x\:=\:199\:n\:m$ and $x\:=\:274\:n\:m$ (ensemble C of 16 red resonance lines with resonance magnetic fields occurring around 1.28 Tesla). In order to see more clearly that those electron spin qubits resonance lines (C) are shifted towards high magnetic field values by the dipolar magnetic field $B_{dip,z}\left(x,\:0\right)$ produced by the ferromagnetic nanostripe and previously calculated, I also added on figure 4a the field sweep spectrum expected for an electron spin qubit created at infinite distance from the ferromagnetic stripe along the x axis (single A red resonance line with resonance magnetic field occurring around 1.215 Tesla) and also the field sweep spectrum expected for an electron spin qubit created at distance $x\:=\:500\:n\:m$ from the ferromagnetic stripe (single B red resonance line with resonance magnetic field occurring around 1.250 Tesla). The most important result shown on figure 4a and highlighted by  figure 4b, is the possibility to design the electron spin based quantum register proposed here such that no spectral overlap occur between the 16 resonances lines of the shifted electron spin qubits and the resonance lines of the two confined spin waves of the ferromagnetic nanostripe occurring at the highest magnetic field values, the so called edge spin wave modes, occurring here at Q band, at 1.227 Tesla and at 1.362 Tesla. This provides a magnetic field interval of 1350 G free of any spin wave resonance. Further assuming an inhomogeneous linewidth below 1 G for each silicon vacancy electron spin qubit in nuclear spin free isotopically purified SiC, one could hope to build identical parallel arrays of 675 electron spin qubits along the x axis. However, state of art pulse EPR spectrometers operating at Q band have a resonator bandwidth of around 1 GHz, corresponding here to a magnetic field interval of around 350 G, in which only around 175 electron spin qubits could be manipulated using selective microwave pulses~\cite{schweiger2001}. It seems thus possible in principle to build a model quantum register of this kind having tens of electrons spins qubits with paramagnetic resonance lines not overlapping the confined spin wave resonance lines, thus avoiding any coherently driven spin decoherence.     

\section{Investigation of the spin qubit decoherence process due to incoherent magnetic fluctuations of the nearby nanoferromagnet}  

Till now, I have not considered the fact that the incoherent thermal excitation of ferromagnetic spin waves can produce some time dependent fluctuating dipolar magnetic field outside the ferromagnetic nanostripe, at the positions where the electron spin qubits are created. Those fluctuations always exist in a ferromagnetic nanostripe at equilibrium at a non zero temperature T. As it is also known from the standard density matrix theory of electron spin decoherence~\cite{tribolletEPJBtheodeco}, any fluctuating magnetic field lead to electron spin decoherence. Thus here, one expects that the fluctuating dipolar magnetic field acting on electron spin qubits will produce a new spin decoherence process, which must be added to the other electron spin decoherence processes, which are intrinsic to the silicon vacancies in the silicon carbide matrix alone. The analysis of this new decoherence process requires the knowledge of the quantum correlation functions of this time dependent fluctuating dipolar magnetic field outside the ferromagnetic nanostripe~\cite{tribolletEPJBtheodeco}. This is a complicated problem not addressed to date. 
\begin{figure} [ht]
\centering \includegraphics[width=1.10\textwidth]{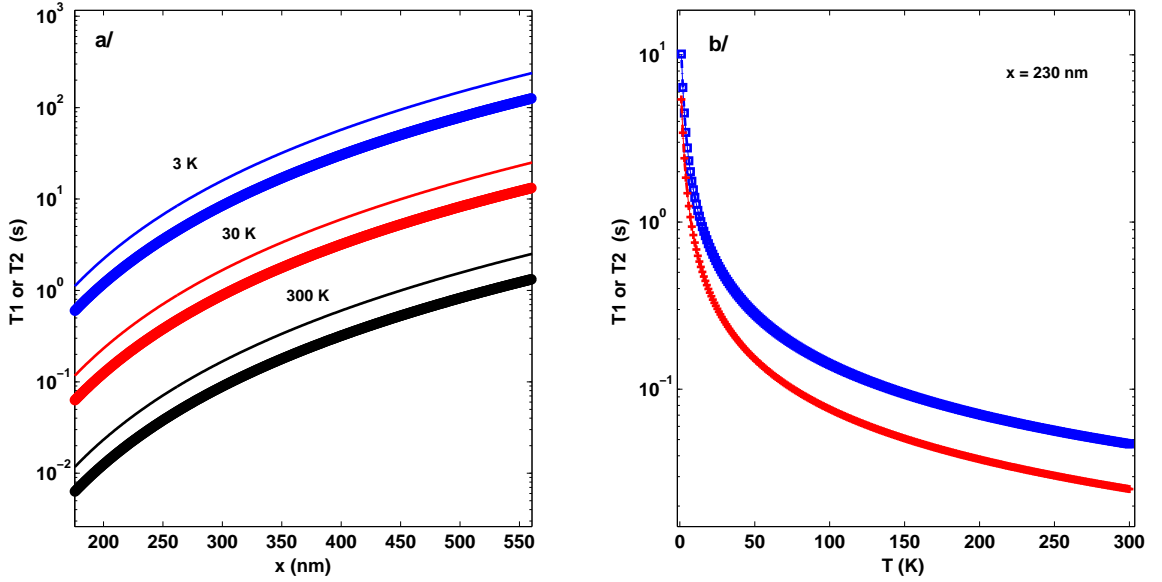} 
\caption{\label{fig_05} Effects of the thermal fluctuations of the dipolar magnetic field produced by the ferromagnetic nanostripe on the electron spin coherence time T2 and on the electron spin longitudinal relaxation time T1 of an electron spin qubit of the nanodevice shown on figure 1. a/ T2 (thin lines) and T1 (thick lines), as a function of the qubit position x for three different temperatures T= 3 K, 30 K, and 300 K. b/ T2 (squarre) and T1 (cross), as a function of the temperature of the nanodevice, assuming an electron spin qubit position $x\:=\:x_{optim}\:=\:230\:n\:m$, where its resonant magnetic field at Q band (34 GHz) is around 1.28 Tesla, and also assuming y=z=0. Permalloy nanostripe dimensions considered: length $L\:=\:100\:\mu\:m$ along the y axis, width $T\:=\:100\:n\:m$ along the x axis, and depth $W\:=\:800\:n\:m$ along the z axis, as shown on figure 1. The estimated T2 and T1 are valid as long as the shifted spin qubit resonance line do not overlap any confined spin wave resonance line, as it is requiered to use the nanodevice shown on figure 1. It was also assumed that the ferromagnetic resonance linewidth of the permalloy nanostripe is equal to 3 $\mu\!e\!V$ at Q band.}

\end{figure}
Here I present the results of a theoretical study of this problem based on the formalisms of the second quantization of spin waves and of the density matrix (details will be published elsewhere). The theory developed shows that this new electron spin decoherence process depends mainly, on the saturation magnetization of the nanoferromagnet, on the temperature, on the spectral detuning between the paramagnetic resonance of the electron spin qubit and the ferromagnetic resonance of the nanostripe, and also on the distance x of the qubit to the ferromagnetic nanostripe, as long as the qubit has a position assumed far from the edges of the ferromagnetic nanostripe along the y and z axis (y<<L, z<<W and x>T/2). Results of the calculation of the electron spin coherence time $T_{2}$ and of the electron spin longitudinal relaxation time $T_{1}$ of an electron spin qubit of the nanodevice shown on figure 1 and operated at microwave Q band are presented on figure 5. At helium temperature, at microwave Q band, and with an applied magnetic field of around 1.28 Tesla, one finds: $T_{1}\:\left(x\:=\:230\:n\!m,\:2\:K\right)\:=\:3.4\:\:s$ and  $T_{2}\:\left(x\:=\:230\:n\!m,\:2\:K\right)\:=\:6.4\:s$. In those experimental conditions, this new decoherence process for the electron spin qubits should thus be negligible compared to other decoherence processes intrinsic to natural SiC. However, it could also become the dominant decoherence process for electron spin qubits embedded in a nuclear spin free isotopically purified SiC matrix, assuming intrinsic spin coherence time in SiC similar to those found in Si, in the range of few seconds~\cite{Tyryshkin2011}. This suggest that quantum information processing with tens of electrons spins qubits encoded onto silicon vacancies in SiC, coherently manipulated by microwave pulse and submitted to the strong dipolar magnetic field gradient of the permalloy ferromagnetic nanostripe should be possible, at least at cryogenic temperatures. Also, at room temperature one finds $T_{1}\:\left(x\:=\:230\:n\!m\:,\:300\:K\right)\:=\:25\:m\!s$ and   $T_{2}\:\left(x\:=\:230\:n\!m\:,\:300\:K\right)\:=\:47\:m\!s$. Thus, those theoretical estimates, combined with the recent observation of room temperature spin coherence of some silicon vacancies in SiC over $80\:\mu\!s$~\cite{Baranov2012}, suggest that a SiC-Permalloy quantum register such as the one designed here should keep its quantum coherence properties at room temperature at a sufficiently high level to perform fundamental studies of quantum entanglement over tens of dipolar coupled electrons spins qubits of silicon vacancies at room temperature. 

\section{Discussion}

SiC is a very attractive material for the practical implementation with a top-down approach of the general quantum register design proposed here, firstly due to the long intrinsic electron spin coherence time of its paramagnetic defects~\cite{Baranov2012} similar to the one in silicon~\cite{Tyryshkin2011}, secondly due to the possibility to create silicon vacancies arrays in SiC using advanced TEM methods, and thirdly due to the possibilities for optical initialization and optical readout of the spin state of some silicon vacancies in SiC~\cite{Awschalom2013,Baranov2012}. Note that a dysprosium ferromagnetic nanostripe would produce a three times bigger magnetic field gradient than in the case of permalloy. However, one expects with dysprosium an electron spin decoherence rate at least one order of magnitude larger than in the case of permalloy at cryogenic temperature due to its three times larger saturation magnetization and its much larger ferromagnetic resonance linewidth. In addition, Dysprosium is no more ferromagnetic at room temperature, contrary to permalloy, which suppresses any possibility for room temperature quantum information processing. Thus the hybrid paramagnetic-ferromagnetic nanodevice presented here, based on the SiC material widely used for high temperature electronics and on the permalloy material widely used for magnetic elements in  computers , seems to be an excellent choice for quantum information processing with dipolar coupled electron spins selectively and coherently manipulated by resonant microwave pulses.


\section{Competing financial interests}
The author declare that he has no competing financial interests.

\end{document}